\documentclass[twocolumn,superscriptaddress,showpacs,
floatfix,preprintnumbers, nofootinbib]{revtex4}

\usepackage{bm}
\usepackage{mathtext}
\usepackage{amssymb, amsmath}
\usepackage{graphicx}
\usepackage{epsfig}
\usepackage{bm}
\usepackage[usenames,dvipsnames]{xcolor}
\usepackage{color}

\def\mystrut{\vrule width0pt depth2pt height8pt}

\begin{document}

\title{Supernova deleptonization asymmetry:
Impact on self-induced flavor conversion}

\author{Sovan Chakraborty}
\affiliation{Max-Planck-Institut f\"ur Physik
  (Werner-Heisenberg-Institut), F\"ohringer Ring 6, 80805 M\"unchen,
  Germany}

\author{Georg Raffelt}
\affiliation{Max-Planck-Institut f\"ur Physik
  (Werner-Heisenberg-Institut), F\"ohringer Ring 6, 80805 M\"unchen,
  Germany}

\author{Hans-Thomas Janka}
\affiliation{Max-Planck-Institut f\"ur Astrophysik,
  Karl-Schwarzschild-Str.~1, 85748 Garching, Germany}

\author{Bernhard M\"uller}
\affiliation{Monash Centre for Astrophysics, School of Physics and Astronomy, Building
79P, Monash University, Victoria 3800, Australia}

\date{December 3, 2014, revised September 4, 2015}

\begin{abstract}
During the accretion phase of a core-collapse supernova (SN), the
deleptonization flux has recently been found to develop a global
dipole pattern (LESA---Lepton Emission Self-sustained Asymmetry). The
$\nu_e$ number flux $F_{\nu_e}$ is much larger than $F_{\bar\nu_e}$ in
one direction, whereas they are approximately equal, or even
$F_{\nu_e}\alt F_{\bar\nu_e}$, in the opposite direction. We use a
linearized stability analysis in a simplified SN model to study the
impact of the $\nu_e$--$\bar\nu_e$ flux asymmetry on self-induced
neutrino flavor conversion. While a small lepton-number flux
facilitates self-induced flavor conversion, ``multi-angle matter
suppression'' is more effective.  Overall we find that for large
matter densities which are relevant below the shock wave, self-induced
flavor conversion remains suppressed in the LESA context and thus
irrelevant for neutrino-driven explosion dynamics.
\end{abstract}


\maketitle

\section{Introduction}
\label{sec:intro}

The neutrino and antineutrino flux spectra emitted by a core-collapse
supernova (SN) significantly depend on flavor. Therefore, flavor
conversion can strongly modify what neutrinos do after decoupling,
notably in driving the explosion, determining nucleosynthesis yields
in the neutrino-driven wind, and the expected signal in large-scale
detectors \cite{Mirizzi:2015eza}.  Even though neutrino mixing angles
are large, in a dense medium the eigenstates of propagation and those
of interaction are very nearly the same~\cite{Wolfenstein:1977ue}.
Therefore, significant flavor conversion would only occur by the MSW
effect~\cite{Mikheev:1986gs, Dighe:1999bi} at a large distance from
the collapsed core.

This situation can fundamentally change when taking neutrino-neutrino
refraction into account~\cite{Pantaleone:1992eq}. It can lead to self-induced
flavor conversion~\cite{Kostelecky:1993dm, Samuel:1995ri} even when the
matter density is large and therefore the effective mixing angle is small
\cite{Duan:2005cp, Duan:2006an, Duan:2010bg}. The propagation eigenmodes of
the collective neutrino ensemble include run-away solutions in flavor
space~\cite{Banerjee:2011fj}, leading to this phenomenon. One question is if
self-induced flavor conversion would occur in regions below the stalled shock
wave during the SN accretion phase. Flavor conversion could then modify
neutrino energy deposition and impact the explosion dynamics in the framework
of the neutrino-driven mechanism of SN explosion.

While self-induced flavor conversion can occur at much higher density than
MSW conversion, it is still suppressed by the ``multi-angle matter effect''
\cite{EstebanPretel:2008ni}, although the exact conditions where conversions
would occur require a linearized stability analysis or a numerical solution
of the neutrino-flavor equations of motion. Dedicated studies, using
one-dimensional (1D) SN models found that the ``onset radius'' of
self-induced flavor conversion would always lie beyond the shock front where
matter densities are much smaller~\cite{Chakraborty:2011nf,
Chakraborty:2011gd}. Subsequent studies using other SN models found similar
results~\cite{Dasgupta:2011jf, Sarikas:2011am}.

Since that time, our theoretical understanding of both self-induced flavor
conversion and of flavor-dependent SN neutrino emission have both evolved. We
are here especially concerned with developments having to do with spontaneous
breaking of symmetries that were previously taken for granted. Axial symmetry
of the neutrino radiation field around a given radial direction had been
assumed in studies of collective flavor oscillations. However, a new class of
run-away solutions breaks this symmetry spontaneously, allowing self-induced
flavor conversion for both neutrino mass orderings~\cite{Raffelt:2013rqa,
Raffelt:2013isa, Mirizzi:2013rla, Mirizzi:2013wda, Hansen:2014paa,
Chakraborty:2014nma}. Moreover, spatial uniformity can also be broken
spontaneously, leading to small-scale instabilities~\cite{Duan:2014gfa,
Mangano:2014zda, Mirizzi:2015fva, Mirizzi:2015hwa}. However, very recently it
was shown that in the SN context, after including multi-angle matter effects,
the uniform mode provides the most sensitive test for
instability~\cite{Chakraborty:2015tfa}. Therefore, in our present study we do
not worry about small-scale instabilities when addressing the impact of an
unusually small lepton-number flux.

The other new finding is that SN neutrino emission in 3D models can
develop a global dipole pattern termed LESA for Lepton Emission
Self-sustained Asymmetry \cite{Tamborra:2014aua,
  Tamborra:2014hga}. While the overall neutrino luminosity remains
nearly spherically symmetric, the deleptonization flux ($\nu_e$ minus
$\bar\nu_e$) develops a strong dipole pattern within the first
$\sim$150~ms after collapse irrespective of other hydrodynamical
instabilities, notably even in the presence of the standing accretion
shock instability (SASI).  More recent 3D models of the Garching group
\cite{Melson:2015tia, Melson:2015spa} as well as the Oak Ridge group
\cite{Lentz:2015nxa} show explosions. While all Garching models
exhibit the LESA effect very clearly and suggest that it persists at
least until some 100~ms after the onset of the
explosion~\cite{Janka:2015}, the 3D explosion model of Oak Ridge shows
a less stationary dipole with an amplitude of just about 10\% of the
monopole over most of the pre-explosion evolution and a more LESA-like
episode with a dipole-to-monopole ratio of $\sim$25\% and stable
direction only in a time interval between 360 and 410~ms after
bounce~\cite{Lentz:2015}.  The reason for these differences between
the Garching and Oak Ridge models is not clear yet, but they may be
connected to different structural or transport conditions in the newly
formed neutron star.  The LESA phenomenon, when it is fully developed,
implies that during the accretion phase, the deleptonization flux
emerges primarily in one hemisphere. In some directions of the
opposite hemisphere, it can be very small and sometimes even slightly
negative.

One may wonder if an unusually small lepton-number flux would facilitate
self-induced flavor conversion, perhaps even below the shock front, and thus
would require a fundamental re-examination of neutrino heating in directions
of small lepton-number flux. For isotropic neutrinos or SN neutrinos in the
single-angle approximation, self-induced flavor conversion is unavoidable in
a $\nu_e$-$\bar\nu_e$ symmetric ensemble~\cite{Hannestad:2006nj, Duan:2007mv,
Raffelt:2007yz, EstebanPretel:2007ec}. On the other hand, multi-angle effects
caused by matter or even by neutrinos themselves can suppress the
instability. The main purpose of our work is a parametric study of the impact
of the $\nu_e$-$\bar\nu_e$ flux asymmetry on self-induced flavor conversion.

We use a linearized stability analysis, where we neglect several issues that
have been discussed in the recent literature. We ignore spin-flip effects
caused by neutrino magnetic dipole moments \cite{Giunti:2014ixa,
deGouvea:2012hg, deGouvea:2013zp} or simply by refraction in inhomogeneous or
anisotropic media \cite{Studenikin:2004bu, Vlasenko:2013fja,
Cirigliano:2014aoa, Vlasenko:2014bva}, and we ignore the role of anomalous
neutrino-antineutrino pair correlations \cite{Volpe:2013uxl,
Vaananen:2013qja, Serreau:2014cfa, Kartavtsev:2015eva}. Realistically, all of
these are probably small effects. It could be more questionable that we also
ignore the ``halo flux'' produced by residual scattering beyond the neutrino
sphere and especially its ``backward going'' component 
\cite{Cherry:2012zw, Cherry:2013mv, Sarikas:2012vb}. Moreover, we do
not worry about the question if our assumed quasi-stationary neutrino source
indeed produces a quasi-stationary solution. A true assessment of these
latter assumptions requires a deeper conceptual development irrespective of
the LESA context.

We begin in Sec.~\ref{sec:flv-evol} with setting up the linear stability
analysis and clarifying our conventions and normalizations of crucial input
parameters. In Sec.~\ref{sec:analysis} we perform the stability analysis for
a simplified setup and we discuss the role of the $\nu_e$-$\bar\nu_e$
asymmetry. We conclude in Sec.~\ref{sec:conclusions} with a summary and
discussion of our findings.

\section{Flavor stability conditions}
\label{sec:flv-evol}

\subsection{Equations of motion}

We set up our stability analysis along the lines of
Ref.~\cite{Raffelt:2013rqa}, slightly adapting some conventions. The
flavor-dependent SN neutrino fluxes depend, at a given radius $r$, on
energy $E$. The neutrino direction of propagation in the transverse
direction is described by the zenith angle $\theta$ relative to the
radial direction and by the azimuth angle $\varphi$. We assume that
the fluxes do not depend on coordinates in the transverse
direction. The impact of small-scale variations in the transverse
direction was recently studied in Ref.~\cite{Chakraborty:2015tfa} with
the result that, for a stability analysis, the largest-scale modes are
most relevant. We use a very basic ``neutrino bulb model'' to describe
the initial fluxes, i.e., we assume that neutrinos are produced at a
spherical surface with radius $R$ and emitted isotropically
(blackbody like) into the outside half-space.  It is convenient to
describe the zenith direction with a fixed label $u=\sin^2\theta_{R}$,
where $\theta_R$ is the zenith angle at radius $R$. Our blackbody
like emission model corresponds to a uniform distribution on the
interval $0\leq u\leq 1$.

We stress that our reference radius $R$ is not a physical quantity but a
quantity of mathematical convenience. It need not coincide with the
neutrino sphere $R_\nu$ where typical neutrinos decouple and which is
the physical emission region.  Neutrinos emerging at $R_\nu$ tend to
show a forward-peaked zenith-angle distribution. The approximate width
of this distribution provides us with the approximate radius $R$ of a
hypothetical blackbody surface that would produce a flux with similar
angular divergence.  Our main point of convenience is to use a $u$
distribution on the unit interval and we define $R$ accordingly.

Notice that we do not integrate the flavor evolution equations, but
the philosophy of the stability analysis is to assume that nothing
happens until our chosen test radius $r$, i.e., neutrinos have
remained in flavor eigenstates which coincide with propagation
eigenstates in a dense medium. Therefore, the exact path taken before
reaching the radius $r$ is irrelevant. We only need to know the
flavor-dependent number fluxes and their angular divergence at radius
$r$.

We combine the flavor-dependent fluxes, assumed to be stationary, into a
matrix ${\sf F}_{r,E,u,\varphi}$ which depends on the variables denoted by
subscripts. Matrices in flavor space are written as capital sans-serif
letters. The diagonal entries are the $4\pi$ equivalent ordinary neutrino
fluxes $F_{\alpha}$ for species $\alpha$. Notice that $F_{\alpha}$ is
differential with regard to energy $E$, i.e., it is a flux spectrum,
and it is also differential with regard to the zenith-angle variable $u$ and
the azimuth angle $\varphi$. The off-diagonal elements of ${\sf F}$ contain
flavor coherence information.

We use negative energies $E$ and negative flux
values $F_{\alpha}$ to denote antineutrinos. This ``flavor isospin
convention'' considerably simplifies the analysis because we do not need to
distinguish explicitly between neutrinos and antineutrinos in the
equations. However, this notation can also be confusing.
$F_{\nu_e}(E)$ is a positive number for
a positive $E$, but a negative number for negative $E$ and then
denotes the antineutrino flux in the sense that $F_{\bar\nu_e}(|E|)=
-F_{\nu_e}(-|E|)$ if we take  $F_{\bar\nu_e}(|E|)$ to be the usual
positive-valued $\bar\nu_e$ flux.

Assuming that the solution does not depend on global direction, the radial
evolution is given by~\cite{Banerjee:2011fj, Raffelt:2013rqa}
\begin{equation}
\textrm{i}\partial_r{\sf F}_{E,u,\varphi}=
[{\sf H}_{E,u,\varphi},{\sf F}_{E,u,\varphi}]\,,
\label{eq:eom1}
\end{equation}
where we have suppressed the index $r$ on all quantities. The Hamiltonian
matrix governing the evolution is \cite{Raffelt:2013rqa}
 \begin{eqnarray}\label{eq:Hamiltonian}
{\sf H}_{E,u,\varphi}&=&\frac{1}{v_{u}} \left(\frac{{\sf M}^2}{2E}
+ \sqrt{2} G_{\rm F} {\sf N}_\ell\right)
 \label{eq:eom2}\\
 &+& \frac{\sqrt{2}G_{\rm F}}{4\pi r^2} \int d \Gamma^{\prime}
 \left(\frac{1-v_{u}v_{u^\prime}-{\bm\beta}_{u}\cdot {\bm\beta}_{u^\prime}}
{v_{u}v_{u^\prime}} \right) {\sf F}'\,,
\nonumber
 \end{eqnarray}
where ${\sf M}^2$ is the neutrino mass-squared matrix, causing vacuum
oscillations. The factor $v_u^{-1}$ arises from projecting the propagation
path onto the radial direction. This factor causes the multi-angle matter
effect: neutrinos traveling in different directions accrue different phases
along the radial direction. The matter term is given by the matrix ${\sf
N}_\ell$ of net charged-lepton densities which is diagonal in the
weak-interaction basis. The third term represents neutrino-neutrino
refraction which is given by the phase-space integral $\int d \Gamma'=
\int_{-\infty}^{+\infty}dE' \int_{0}^{1} du' \int_{0}^{2\pi}d\varphi'$, where
${\sf F'}$ is understood as ${\sf F}_{E',u',\varphi'}$ at radius $r$. The
radial velocity of a given mode is $v_{u} = (1-u R^2/r^2)^{1/2}$
\cite{EstebanPretel:2007ec}. The transverse velocity ${\bm \beta}$ depends on
the azimuth angle $\varphi$. One finds $|{\bm \beta}_{u}|=\sqrt{u}\, R/r$ and
${\bm\beta}_{u}\cdot {\bm\beta}_{u^\prime}=\sqrt{u
u'}(R/r)^2\cos(\varphi-\varphi')$ \cite{Raffelt:2013rqa}.

We restrict our stability analysis to a two-flavor system and express the
flux matrices in the form
\begin{equation}
{\sf F}=\frac{{\rm Tr}\,{\sf F}}{2}
+\frac{F_{\nu_e}^R-F_{\nu_x}^R}{2}\,{\sf S}\,,
\quad
{\sf S}=\left( \begin{array}{cc} s &  S \\
S^{\ast} & -s
\end{array} \right)\,,
\label{eq:pHI}
\end{equation}
where $s$ is real, $S$ complex, and $s^2+|S|^2=1$. The fluxes with
superscript $R$ are the ones at radius $R$ where $s=1$ and $S=0$. Since the
overall neutrino flux is conserved, the trace term is conserved and drops out
of the commutator equation. Because the $F_\alpha$ are $4\pi$ equivalent
fluxes, they do not decrease with distance, only the flavor content evolves.
The flavor evolution is encoded in the normalized matrix ${\sf S}$.

Next we change the energy variable $E$ to the frequency variable $\omega=
\Delta m^2/2E$ which is more convenient in the flavor oscillation context.
Negative $\omega$ values denote antineutrino modes of the neutrino field. We
imagine that self-induced flavor conversion is driven by the atmospheric mass
difference. Our sign convention is such that a positive $\Delta m^2$ refers
to inverted mass ordering. Actually we will always assume $\Delta m^2$ to
be a positive parameter and will explain later on how to implement normal
mass ordering in our equations.

As far as flavor conversion is concerned, what enters is the two-flavor flux
difference $F_{\nu_e}-F_{\nu_x}$, not the individual flavor fluxes.
Therefore, it proves useful to introduce a dimensionless neutrino spectrum in
the form
\begin{equation}\label{eq:gdef}
g=\frac{F_{\nu_e}^R-F_{\nu_x}^R}{\frac{1}{2}
\int d\Gamma \left|F_{\nu_e}^R-F_{\nu_x}^R\right|}\,,
\end{equation}
where $g$ and all fluxes depend on $\omega$, $u$, and $\varphi$, and
\begin{equation}\label{eq:gammadef}
\int d\Gamma= \int_{-\infty}^{+\infty}d\omega \int_{0}^{1} du
\int_{0}^{2\pi}d\varphi\,.
\end{equation}
We recall that antineutrino fluxes in the flavor-isospin convention are
negative and that typically the $\nu_e$ flux is larger than the $\nu_x$ flux,
and similar for antineutrinos. Therefore, in the SN context, $g$ is typically
positive for positive $\omega$ (neutrinos) and negative for negative $\omega$
(antineutrinos), although the opposite sign can appear for some range of
$\omega$, corresponding to a ``crossed over'' number flux between $\nu_e$ and
$\nu_x$.

Our normalization differs from our previous studies where we often used
\smash{$\int_{-\infty}^{0}d\omega \int_{0}^{1} du
\int_{0}^{2\pi}d\varphi\,\left|F_{\bar\nu_e}^R-F_{\bar\nu_x}^R\right|$} in the
denominator, i.e., the total number flux of $\bar\nu_e$ minus that of
$\bar\nu_x$. However, in the maximal lepton-flux direction of our 3D models,
this quantity becomes very small or even vanishes and thus can not be used to
normalize other fluxes. Moreover, our new definition is more symmetric and
more intuitive.

With the spectrum thus normalized, we define the \hbox{$\nu_e$-$\bar\nu_e$}
asymmetry parameter as
\begin{equation}\label{eq:eps-from-g}
\epsilon=\int d\Gamma\,g(\omega,u,\varphi)
=\frac{\int d\Gamma\,\left(F_{\nu_e}^R-F_{\nu_x}^R\right)}{\frac{1}{2}
\int d\Gamma \left|F_{\nu_e}^R-F_{\nu_x}^R\right|}\,.
\end{equation}
If we denote with $N_\alpha$ the total positive number flux for neutrinos or
antineutrinos $\alpha$, the asymmetry parameter corresponds to
\begin{equation}\label{eq:eps-definition}
\epsilon=2\,\frac{(N_{\nu_e}-N_{\bar\nu_e})-(N_{\nu_x}-N_{\bar\nu_x})}
{(N_{\nu_e}+N_{\bar\nu_e})-(N_{\nu_x}+N_{\bar\nu_x})}\,.
\end{equation}
In the SN context we have approximately $N_{\nu_x}=N_{\bar\nu_x}$, further
simplifying this expression. The normalization in our previous papers was
$\epsilon=(N_{\nu_e}-N_{\bar\nu_e})/(N_{\bar\nu_e}-N_{\bar\nu_x})$ in this
notation.

The main point of our study is to investigate how flavor stability depends on
the asymmetry parameter $\epsilon$. In the LESA context, we have a strong
lepton-number flux dipole, whereas the $\nu_e$ plus $\bar\nu_e$ and the
$\nu_x$ plus $\bar\nu_x$ fluxes remain nearly $4\pi$ symmetric, i.e., the
denominator in our definition of $\epsilon$ is nearly independent of
direction.

\subsection{Linearization}

We finally linearize the equations of motion in two ways. We assume
$|S|\ll 1$ as we want to investigate exponentially growing modes in
$S$, implying $s=1$ to linear order. Furthermore, we use the
large-distance approximation, $r\gg R$, leading
to~\cite{Raffelt:2013rqa},
\begin{eqnarray}\label{eq:linearizedEOM}
\kern-1em
\textrm{i}\partial_r S &=& [\omega + u(\lambda +\epsilon \mu)] S  \nonumber \\
&-& \mu \int  d \Gamma^\prime [u + u^\prime - 2\sqrt{u u^\prime} \cos
(\varphi-\varphi^\prime)]g^\prime S^\prime
\label{eq:lin} \,,
\end{eqnarray}
where $S$ depends on $\omega$, $u$ and $\varphi$, whereas $S'$ and $g'$
depend on $\omega'$, $u'$ and $\varphi'$. We then seek eigenvalues
$\Omega=\gamma+i\kappa$ for solutions of the form
$S(r,\omega,u,\varphi)=Q_{\omega,u,\varphi}\,e^{-i\Omega r}$, where a
positive imaginary part $\kappa$ reflects unstable solutions.

The effective multi-angle strength of the neutrino-neutrino and
neutrino-matter interaction are described by the parameters
\begin{subequations}\label{eq:mu-lambda-def}
\begin{eqnarray}
\mu &=& \sqrt{2}G_{\rm F}\,
\frac{(N_{\nu_e}+N_{\bar\nu_e})-(N_{\nu_x}+N_{\bar\nu_x})}{8 \pi r^2}\,
\frac{R^2}{2 r^2}\,,
\label{eq:mu}\\
\lambda &=& \sqrt{2} G_{\rm F}\, n_e \frac{R^2}{2 r^2}\,,
\label{eq:lambda}
\end{eqnarray}
\end{subequations}
where $n_e$ is the net electron density ($e^-$ minus $e^+$). On the
other hand, $n_{\nu_e}=N_{\nu_e}/(4\pi r^2)$ is the $\nu_e$ density at
radius $r$ and $n_{\bar\nu_e}=N_{\bar\nu_e}/(4\pi r^2)$ is the
$\bar\nu_e$ number density. Therefore, in the definition of $\mu$,
besides subtracting the $\nu_x$ and $\bar\nu_x$ densities, we use the
sum $(n_{\nu_e}+n_{\bar\nu_e})/2$ to describe the neutrino matter
effect. Our definition of $\mu$ differs somewhat from the previous
literature, corresponding to our modified definition of $g$. What
enters in the equations is the combination
$\mu\,g_{\omega,u,\varphi}$, which is independent of chosen
normalizations of either quantity.

In the LESA context, the total number fluxes $N_{\nu_e}+N_{\bar\nu_e}$
and $N_{\nu_x}+N_{\bar\nu_x}$ are almost $4\pi$ symmetric. Therefore,
our new normalization of $\mu$ implies that it is nearly $4\pi$
symmetric in our models.

Notice also that $\mu$ explicitly shows the usual $r^{-4}$ variation
with distance. Therefore, we will write
\begin{equation}\label{eq:mur}
\mu=\mu_R\,(R/r)^4\,.
\end{equation}
In our explicit analysis we will use $\mu_R=2\times10^5~{\rm km}^{-1}$
and $R=15$~km, whereas the physical neutrino sphere is approximately
at $R_\nu=30$~km. Either $\mu$ or $r$ can be used as a measure of
distance from the SN core.

\subsection{Axially symmetric neutrino emission}

We assume axially symmetric neutrino emission, implying
$g(\omega,u,\varphi)\to g(\omega,u)/2\pi$, but not necessarily an
axially symmetric solution. In the LESA context, the local neutrino
radiation field at some distance $r$ and some specific direction is
not axially symmetric.  However, if we assume LESA to represent an
exact dipole, the system shows global axial symmetry relative to the
dipole direction. Our schematic neutrino distributions always refer to
the direction of minimal or maximal lepton-number emission, i.e., to
the positive or negative LESA dipole directions. In these extreme
lepton-asymmetry directions the fluxes can be assumed to be locally
axially symmetric and in this sense axial symmetry pertains to the
emitted fluxes.  We do not think that our overall conclusions are
strongly affected by this simplification.

As derived in Ref.~\cite{Raffelt:2013rqa}, in the axially symmetric case one
finds the eigenvalue equations
\begin{equation}\label{eq:stabilityconidtion}
(I_1-1)^2=I_0I_2
\quad\hbox{or}\quad
I_1=-1\,,
\end{equation}
where
\begin{equation}\label{eq:In-definition}
I_n=\mu\int d\omega\,du\,\frac{u^n\,g(\omega,u)}{\omega+u(\lambda+
\epsilon\,\mu)-\Omega}\,.
\end{equation}
These eigenvalue equations are to be solved to find the eigenfrequencies
$\Omega$ and the range of parameters where these have an imaginary part,
signifying instabilities with regard to self-induced flavor conversion.

Notice that the first block in Eq.~(\ref{eq:stabilityconidtion})
provides solutions for the axially symmetric case, the so-called
bimodal instability (for inverted mass ordering) and the
multi-zenith-angle (MZA) instability (for normal mass ordering). The
second block provides the instability where axial symmetry is
spontaneously broken, the multi-azimuth-angle (MAA) instability which
appears only for normal mass ordering.

\subsection{Switching neutrino mass ordering}

Our equations are written for inverted neutrino mass ordering (IO),
whereas normal ordering (NO) corresponds to a negative $\Delta
m^2$. However, we prefer to keep $\Delta m^2$ a positive parameter so
that negative $\omega=\Delta m^2/2E$ continues to denote
antineutrinos. We can account for the mass ordering by introducing a
parameter $h=\pm1$ in front of the vacuum oscillation term ${\sf
  M}^2/2E$ in the Hamiltonian matrix Eq.~(\ref{eq:Hamiltonian}). This
parameter carries through and finally appears in the first line of
Eq.~(\ref{eq:linearizedEOM}) in front of the term $\omega$. We can
then, for NO, multiply this equation with $-1$, restoring the original
$\omega$ term, changing the sign of $\mu$ and $\lambda$, and the
l.h.s.\ of the equation. For flavor oscillations, it is irrelevant if
neutrinos oscillate ``clock-wise'' or ``counter-clockwise'' in flavor
space, i.e., the sign change on the l.h.s.\ is not important to find
the eigenvalues. In other words, normal mass ordering is covered by
\begin{equation}
\hbox{IO $\to$ NO:}\quad
\mu\to-\mu \quad\hbox{and}\quad \lambda\to-\lambda\,.
\end{equation}
Therefore, we should solve the eigenvalue equation for
$-\infty<\mu,\lambda<+\infty$. The first quadrant, where both parameters are
positive, corresponds to IO, the third quadrant, where both parameters are
negative, to NO.

\subsection{Switching the \boldmath{$\nu_e$}-\boldmath{$\bar\nu_e$} asymmetry}

We are mainly concerned with the role of the asymmetry parameter
$\epsilon$. Of course, there are many ways one can modify the spectrum
$g(\omega,u)$ to achieve a different asymmetry $\epsilon$. However, as a
simple case we could imagine to switch the role of $\nu_e$ and
$\bar\nu_e$, leaving all else unchanged. This modification corresponds to
\begin{equation}\label{eq:eps-switch}
g(\omega,u)\to -g(-\omega,u)\,,
\end{equation}
which implies $\epsilon\to-\epsilon$. In other words, we consider a
new spectrum $\tilde g(\omega,u)=-g(-\omega,u)$. Following the
propagation of signs in the equations, one finds that this new system
is equivalent to the original one with the sign change
$\lambda\to-\lambda$, whereas $\mu$ remains unchanged.  More
precisely, if instead of switching the sign of $\epsilon$ we switch
the sign of $\lambda$, the eigenvalue will also change sign, i.e.,
$\Omega\to-\Omega$. Whenever $\Omega$ has an imaginary part, there
exists also the complex conjugate solution, i.e., an exponentially
growing and an exponentially damped solution. Therefore, it is
irrelevant if we find $\Omega$ or $-\Omega$ because the growth
rate is the same in both cases.

Finally, then, for a given spectrum $g(\omega,u)$ and studying the stability
condition in the full parameter range $-\infty<\mu,\lambda<+\infty$, all four
quadrants have a physical interpretation according to
\begin{equation}
\begin{matrix}
&\mu<0&\mu>0\\[1ex]
\lambda>0&\fbox{\mystrut NO, $\epsilon<0$}&\fbox{\mystrut IO, $\epsilon>0$}\\[1ex]
\lambda<0&\fbox{\mystrut NO, $\epsilon>0$}&\fbox{\mystrut IO, $\epsilon<0$}\\
\end{matrix}
\end{equation}
where the sign change in $\epsilon$ is understood in the spirit of
Eq.~(\ref{eq:eps-switch}).

\section{Stability analysis}
\label{sec:analysis}

\subsection{Simplified spectrum}

As a next step we perform the stability analysis for a simplified spectrum.
We are not concerned with an exact numerical result for a specific SN model,
but rather we wish to understand the qualitative impact of a modified
$\nu_e$-$\bar\nu_e$ asymmetry when all else is kept fixed. Therefore, we
construct the simplest possible toy model which allows us to develop an
understanding of this question.

For the zenith-angle distribution we assume black-body like emission from the
neutrino sphere, i.e., we assume a uniform distribution of the $u$-variable
on the interval $0\leq u \leq 1$, together with a suitable reference
radius $R$ as explained earlier. For the relationship between test
radius $r$ and effective neutrino interaction energy $\mu$ we use the
relationship shown in and around Eq.~(\ref{eq:mur}).

Our choices are motivated by the 3D models that have led to the
discovery of the LESA phenomenon \cite{Tamborra:2014aua, Tamborra:2014hga}.
Specifically, we use the model with
$11.2\,M_{\odot}$ progenitor mass as a benchmark for our study. During
the accretion phase, this model shows large-scale convective overturn,
but it does not develop the standing accretion shock instability
(SASI), in contrast to the models with larger progenitor masses. After
about 150~ms post bounce (p.b.), this and the other models develop a
large-scale anisotropy of lepton-number emission.

\begin{figure}
\begin{center}
\includegraphics[angle=0,width=0.40\textwidth]{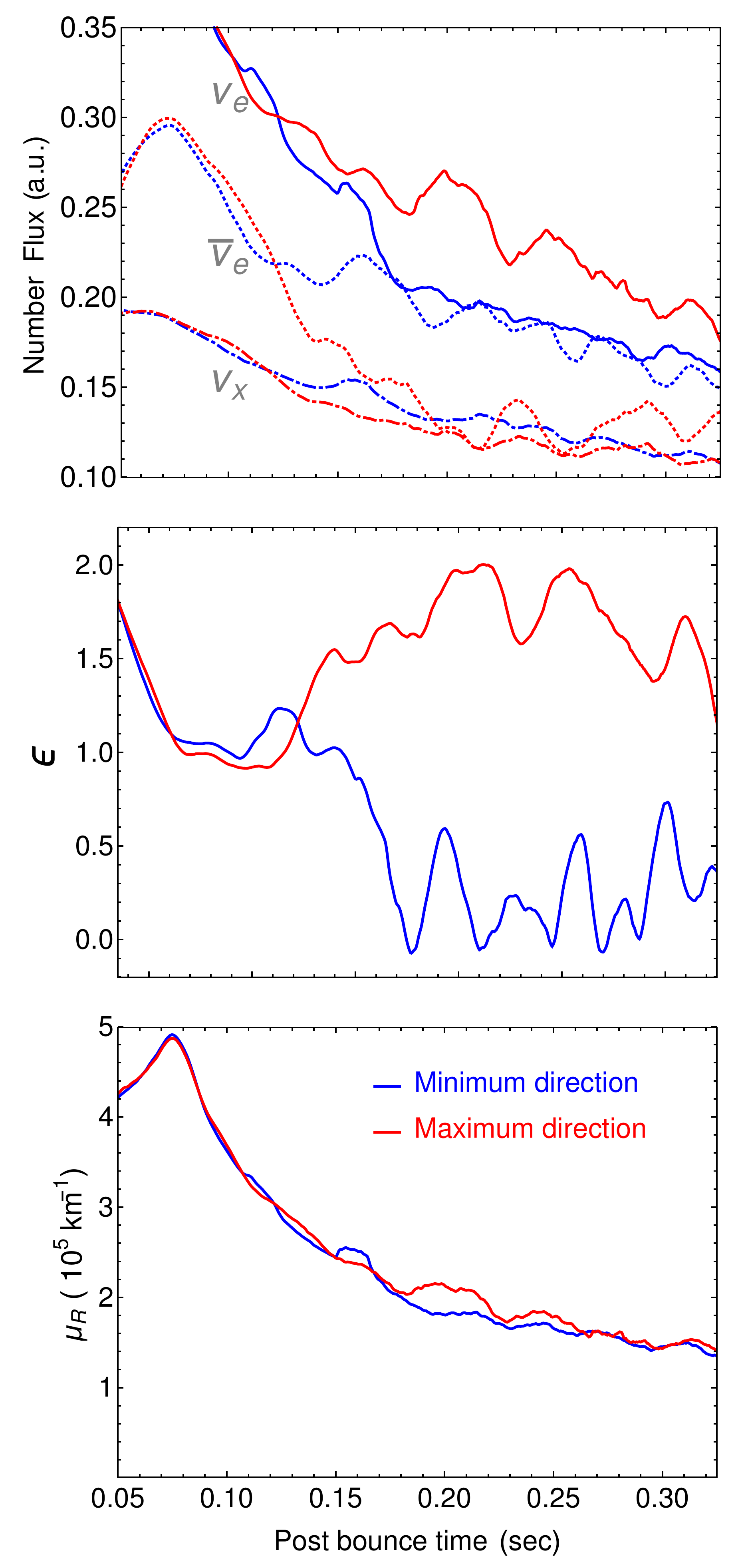}
\end{center}
\caption{Physical characteristics of our benchmark SN model in the
  minimum (blue lines) and maximum (red lines) lepton-flux
  directions. The data have been smoothed (running averages) over
  approximately 20~ms.  {\it Top:}~Number fluxes of the species
  $\nu_e$, $\bar\nu_e$ and $\nu_x$.  The $\nu_x$ flux (representing
  any of $\nu_\mu$, $\bar\nu_\mu$, $\nu_\tau$ and $\bar\nu_\tau$) is
  similar in both directions. The $\bar\nu_e$ flux develops a large
  asymmetry at around 150~ms p.b. In the direction of minimal lepton
  number flux, the $\bar\nu_e$ flux is as large as the $\nu_e$ flux,
  in the opposite direction it is as small as the $\nu_x$ flux.  {\it
    Middle:} Asymmetry parameter $\epsilon$ of the lepton-number flux
  as defined in Eq.~(\ref{eq:eps-definition}).  {\it Bottom:}
  Effective neutrino-neutrino interaction strength $\mu_R$ at the
  reference radius $R=15$~km, see Eq.~(\ref{eq:mur}).}
 \label{fig-epsilon}
\end{figure}

We illustrate this behavior in Fig.~\ref{fig-epsilon} where we show in
the top panel the flavor-dependent neutrino number fluxes in two
opposite directions, roughly corresponding to the directions of
maximal and minimal lepton-number flux, respectively. The $\nu_x$
fluxes in the two opposite directions are similar. However, in the
minimum direction, the $\bar\nu_e$ flux is similar to the $\nu_e$
flux, corresponding to a small lepton asymmetry. In the maximum
direction, the $\bar\nu_e$ and $\nu_x$ fluxes are similar and much
smaller than the $\nu_e$ flux.

From these number fluxes we can derive the asymmetry parameter
$\epsilon$ which we show as a function of time for our two extreme
directions in the middle panel of Fig.~\ref{fig-epsilon}.  When LESA
is fully developed, the $\nu_e$ and $\bar\nu_e$ fluxes are
approximately equal in the direction of minimum lepton-number flux so
that $\epsilon\approx 0$. In the maximum lepton number flux direction,
the ${\bar\nu_e}$ and ${\bar\nu_x}$ fluxes are almost exactly
equal. In our formulation, this situation corresponds to
$\epsilon=2$. In this extreme case, self-induced flavor conversion is
not possible because $N_{\nu_e}-N_{\nu_x}$ cannot be swapped with
$N_{\bar\nu_e}-N_{\bar\nu_x}$ if the latter is zero. Therefore, a
stability analysis for $\epsilon=2$ is moot, but we will consider
a case with  $\epsilon=1.9$.
As an intermediate benchmark case we consider a situation where
$N_{\bar\nu_e}-N_{\bar\nu_x}$ is half of $N_{\nu_e}-N_{\nu_x}$,
corresponding to $\epsilon=2/3$.

Another crucial parameter is the effective neutrino-neutrino
interaction strength defined in Eq.~(\ref{eq:mu}).  We show $\mu_R$ as
a function of time for our model in the bottom panel of
Fig.~\ref{fig-epsilon}.  A typical value is a few $10^5~{\rm
  km}^{-1}$. We will specifically use the value
$\mu_R=2\times10^5~{\rm km}^{-1}$ which is typical for the period
after 150~ms when the LESA dipole is fully developed.  Notice that the
choice of $\mu_R$ is not relevant for the stability analysis per se,
but only to establish a relationship between physical radius $r$ and
corresponding $\mu$-value.

From past experience we know that the exact neutrino energy
distribution tends to be relatively insignificant.  We have explicitly
checked this point by using top-hat shaped, Maxwell-Boltzmann, and
more general Gamma distributions. The locus of the instability region
in the $\mu$--$\lambda$--plane is primarily determined by the average
oscillation frequency $\langle\omega\rangle$.  Because we
study the differential effect caused by modified $\epsilon$ values it
is most transparent to use the simplest possible model which exhibits
these effects.  Therefore, we describe all $\nu_e$ and $\bar\nu_e$ by
a single energy, i.e., the system is described by a vacuum oscillation
frequency $\omega_{\nu}$ and one for $\omega_{\bar\nu}$.  Moreover, we
can always go to a rotating coordinate frame in flavor space such that
effectively $\omega_0\equiv \omega_\nu=-\omega_{\bar\nu}$. Therefore,
our simplified model is described by a spectrum
$g(\omega,u)=h_\epsilon(\omega)$ which does not depend on $u$.  Here,
\begin{equation}\label{eq:monospectrum}
h_\epsilon(\omega)=\left(1+\frac{\epsilon}{2}\right)\,\delta(\omega-\omega_0)
-\left(1-\frac{\epsilon}{2}\right)\,\delta(\omega+\omega_0)\,,
\end{equation}
where the first term is for neutrinos, the second one for antineutrinos.

The abstract stability analysis does not depend on the numerical
choice of $\omega_0$ in the sense that all other quantities of dimension
``frequency,'' i.e., $\lambda$, $\mu$ and $\kappa$ are expressed in units of
$\omega_0$. The chosen value only matters when translating the stability
region in the parameter space of $\mu$ and $\lambda$ to physical SN parameters.
Specifically we use $E = 12$~MeV as our single energy, corresponding
to $\omega_0= \Delta m_{\rm atm}^2/2E = 0.51~{\rm km}^{-1}$. Other choices would slightly
shift the SN density profile relative to the instability region in our final
plots Figs.~\ref{fig:stab-IO} and~\ref{fig:stab-NO}.

The main practical reason for using a single-energy
neutrino spectrum is that it allows a simple analytic integration of
the integrals Eq.~(\ref{eq:In-definition}). Avoiding a numerical integration
considerably accelerates the numerical search for the eigenvalues. The
imaginary part of $\Omega$, the growth rate $\kappa$, is always of order
$\omega_0$, whereas in our range of interest, $\mu$ and $\lambda$ are up to
$10^6$ in these units and the real part of $\Omega$ can also take on such
large values. Therefore, finding the eigenvalues can be numerically
challenging.

\begin{figure*}
\includegraphics[angle=0,width=0.806\textwidth]{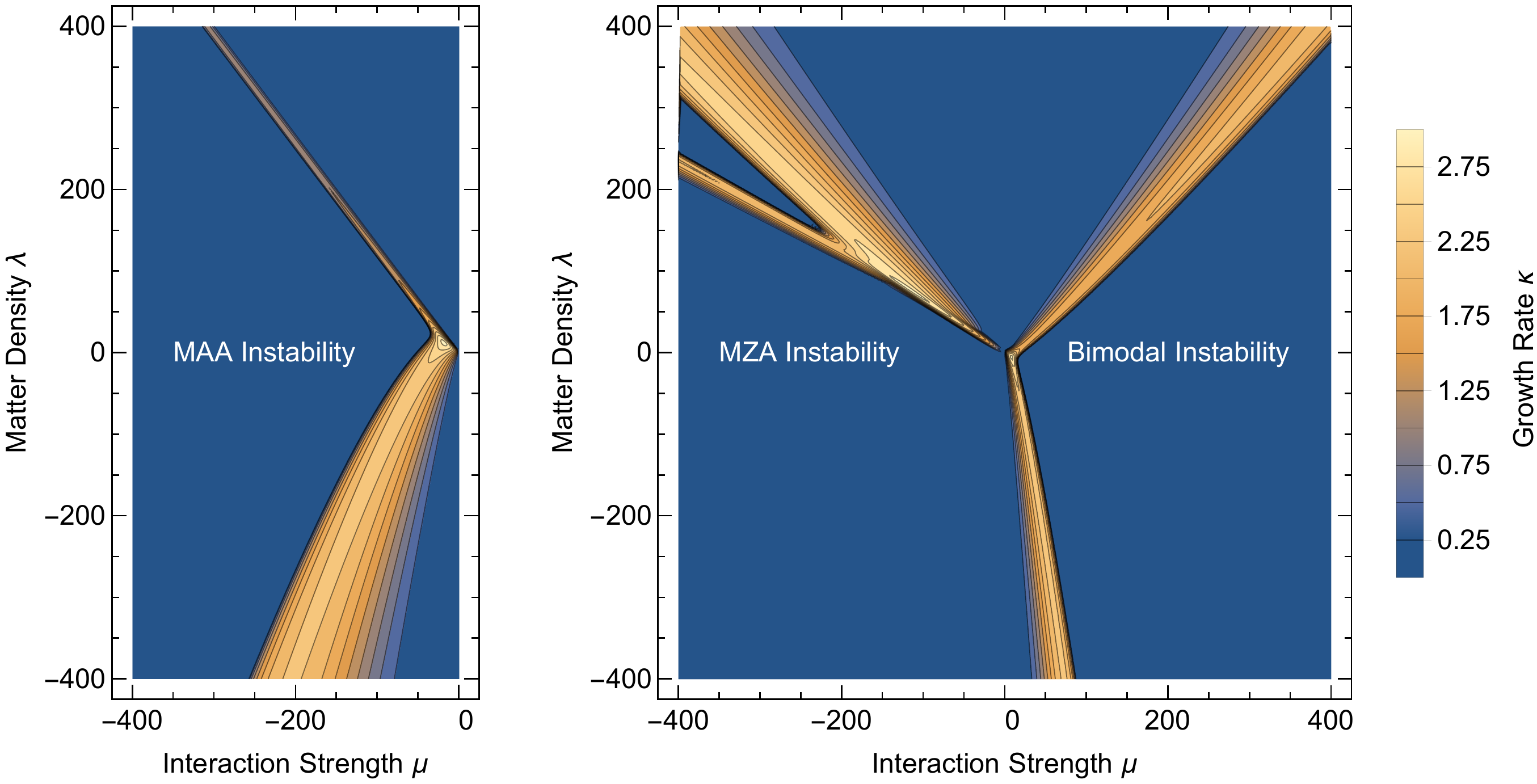}
\caption{Contour plot of the instability growth rate $\kappa$ for the
simplified SN model described in the text. The effective neutrino-neutrino
interaction strength $\mu$ and the matter effect $\lambda$ were
defined in Eq.~(\ref{eq:mu-lambda-def}). Notice that $\kappa$, $\lambda$ and
$\mu$ are given in units of the vacuum oscillation frequency $\omega_0$.
The upper right quadrant ($\mu,\lambda>0$) of the right panel corresponds to inverted mass
ordering (IO) and the asymmetry parameter $\epsilon=+\frac{2}{3}$, shows the
traditional bimodal instability. The lower left quadrant ($\mu,\lambda<0$),
corresponding to normal mass ordering (NO) and \smash{$\epsilon=+\frac{2}{3}$}, shows
only the MAA instability. The upper left and lower right quadrants,
corresponding to opposite signs of $\mu$ and $\lambda$, represent the
\smash{$\epsilon=-\frac{2}{3}$} cases. The right panel (bimodal and MZA
instabilities) arises from the the first block in the eigenvalue equation
(\ref{eq:stabilityconidtion}). The left panel arises from the second block,
deriving from solutions which break axial symmetry (MAA instability).\label{fig:contour1}}
\vskip15pt
\includegraphics[angle=0,width=0.806\textwidth]{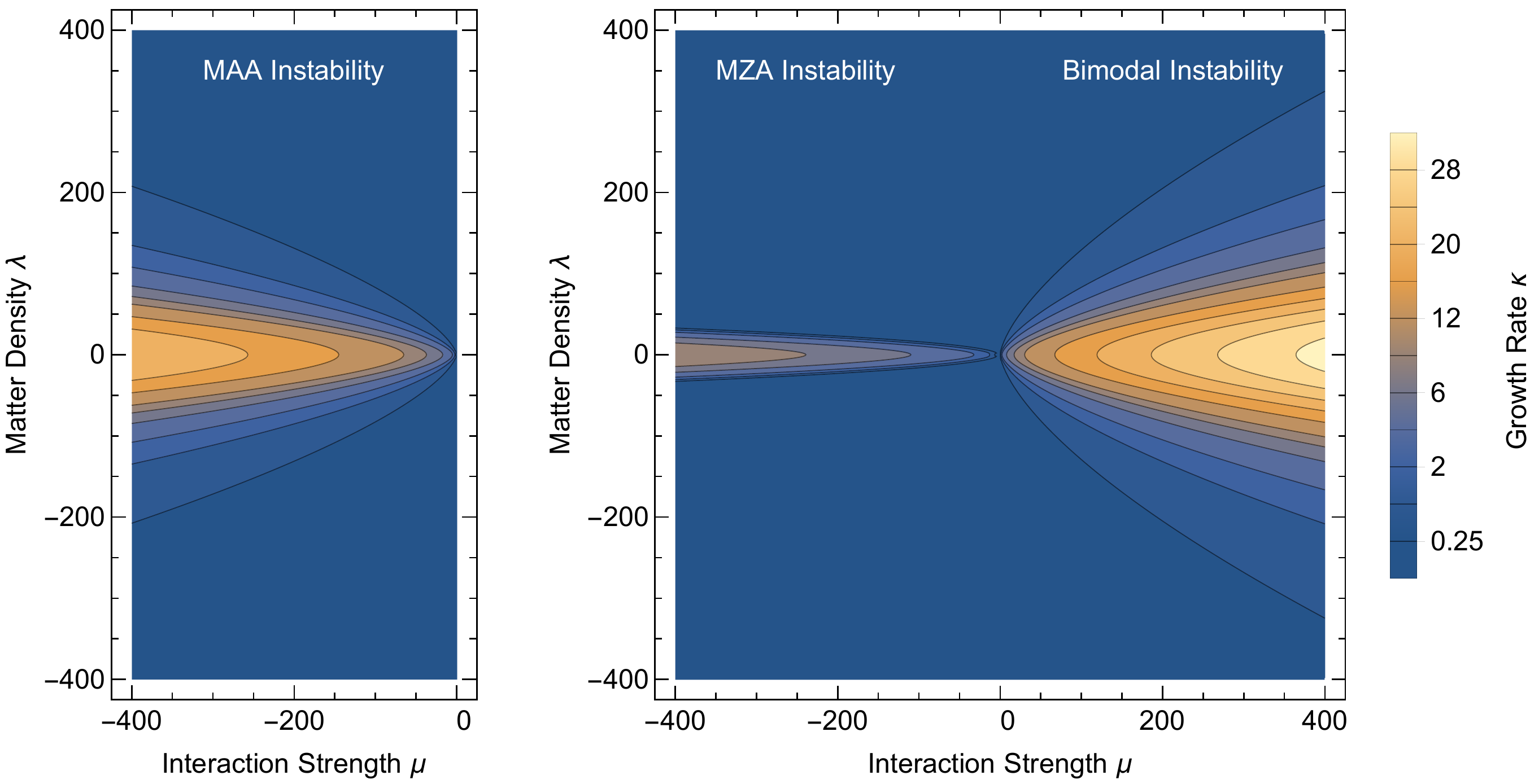}
\caption{Same as Fig.~\ref{fig:contour1} for $\epsilon=0$. Notice the
expanded scale of $\kappa$ values. In both figures, the lowest contour is for
$\kappa=0.25\,\omega_0$.\label{fig:contour2}}
\end{figure*}

\subsection{Structure of the instability regions}

We provide the explicit eigenvalue equations for our simplified
spectrum in Appendix~\ref{sec:explicit}. For general values of $\mu$,
$\lambda$ and $\epsilon$, the eigenvalue $\Omega$ and its imaginary
part must be found numerically. The structure of typical unstable
solutions is best explained using a simple example with
$\epsilon=\frac{2}{3}$, corresponding to twice as many $\nu_e$ than
$\bar\nu_e$ after subtracting the $\nu_x$ and $\bar\nu_x$ fluxes. We
first consider the first block in Eq.~(\ref{eq:stabilityconidtion})
which derives for those solutions which preserve axial symmetry around
a given radial direction. In Fig.~\ref{fig:contour1} we show the
growth rate $\kappa$ for unstable solutions in the parameter space
consisting of the neutrino-neutrino interaction strength $\mu$ and the
multi-angle matter effect $\lambda$ as defined in
Eq.~(\ref{eq:mu-lambda-def}). The upper right quadrant
($\mu,\lambda>0$) of the right panel corresponds to IO and shows the traditional bimodal
instability region. For a given neutrino density $\mu$, the system is
stable if the matter density is either too small or too large. The
same effect can be described as a shift of the unstable $\mu$ range to
larger $\mu$ values for increasing $\lambda$. This effect has been
termed ``multi-angle matter suppression'' of the instability, although
the instability is not suppressed, it is shifted to larger neutrino
densities. However, we can say that it is suppressed relative to a
specific locus in the $\mu$-$\lambda$ plane which corresponds to the
SN density profile.

In the lower left quadrant ($\mu,\lambda<0$) of the right panel, corresponding with our
conventions to NO, there is no instability. Physically, of course,
$\mu$ and $\lambda$ are positive---changing their sign allows us to
show the NO case in the same plot as explained earlier.

When we include axial symmetry breaking a new set of solutions arises,
deriving from the second block in Eq.~(\ref{eq:stabilityconidtion}),
and we find the solutions shown in the left panel. This is the
multi-azimuth angle (MAA) instability discussed in the previous
literature. Overall, then, in IO we find the traditional bimodal
instability, in NO the more recent MAA instability.

The upper left and lower right quadrants ($\mu$ and $\lambda$ have opposite
signs) are unphysical in the sense that these quadrants require an excess
$\bar\nu_e$ flux. In the SN context, the collapsed material is lepton rich,
implying that both the electron and neutrino densities are dominated by
particles, not antiparticles, and thus that $\lambda$ and $\mu$ have the same
sign. However, in the LESA context it can happen that in the direction of
minimal lepton-number flux the $\bar\nu_e$ flux actually dominates. In this
sense, a small negative $\epsilon$ is not entirely hypothetical, although a
large negative value such as $\epsilon=-\frac{2}{3}$ would be unrealistic.
Therefore, the upper left and lower right quadrants in
Fig.~\ref{fig:contour1} are primarily shown for mathematical completeness. In
the lower right quadrant, corresponding to IO with $\epsilon=-\frac{2}{3}$,
the instability range is rather insensitive to $\lambda$. In the upper left
quadrant, corresponding to NO with $\epsilon=-\frac{2}{3}$, we find a total
of three instabilities.

From a mathematical perspective, $\bar\lambda=\lambda+\epsilon\mu$ and
$\mu$ is a more natural pair of parameters. Notice that in the eigenvalue
equations, the matter effect enters in the form
$\bar\lambda=\lambda+\epsilon\mu$, where $\epsilon \mu$ represents the matter
effect caused by neutrinos on each other. In the $\mu$-$\bar\lambda$
parameter space, the two MZA instabilities in the right panel of
Fig.~\ref{fig:contour1} appear in different quadrants (cf.\ Fig.~9 of
Ref.~\cite{Chakraborty:2015tfa}). In other words, the line
$\bar\lambda=\lambda+\epsilon\mu=0$, equivalent to $\lambda=-\epsilon\mu$,
cuts between the two MZA instabilities.

As a second case we show in Fig.~\ref{fig:contour2} a similar contour plot
for the fully symmetric case $\epsilon=0$. As expected, the plot is now
symmetric under the exchange $\lambda\to-\lambda$. For $\epsilon=0$ and in
the absence of matter (on the line $\lambda=0$), the growth rate becomes
arbitrarily large for $\mu\to\infty$. In this symmetric case without matter,
there is no ``sleeping top regime,'' i.e., the system is unstable for any
$\mu$ above a very small threshold. Notice that the eigenvalue equation can
be solved analytically for $\bar\lambda=0$ and any value of $\epsilon$ as shown
in Appendix~\ref{sec:explicit}.

\begin{figure*}
\includegraphics[angle=0,width=1.0\textwidth]{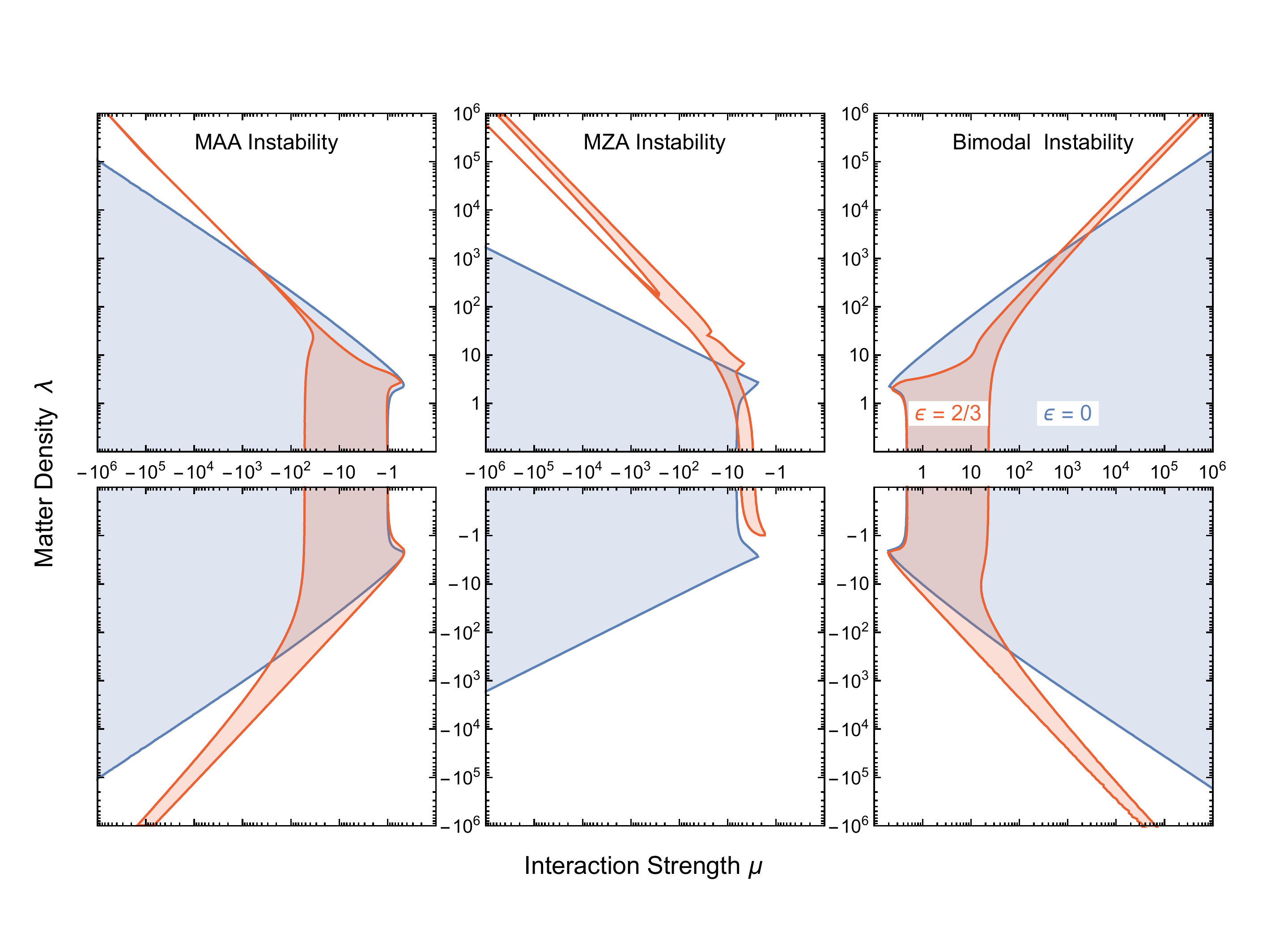}
\caption{Instability footprints corresponding to the contour plots of
Fig.~\ref{fig:contour1} (red for $\epsilon=2/3$) and of
Fig.~\ref{fig:contour2} (blue for $\epsilon=0$), where both $\mu$ and $\lambda$
are measured in units of the vacuum oscillation frequency $\omega_0$.
The limiting growth
rate to construct the footprints is $\kappa=0.01\,\omega_0$.
For inverted mass ordering, the bimodal instability applies (upper right panel).
For normal mass ordering, the lower MZA and MAA panels apply, but in practice
only the lower left panel (MAA) is relevant because the MAA footprint is
closer to the SN density profile than the MZA footprint.
\label{fig:footsix}}
\end{figure*}

\subsection{Instability footprints}

We are here not particularly interested in the details of the growth rates of
the various instabilities, but rather, if neutrinos streaming from a SN core
along a given density profile will encounter any instability. In this sense
it is enough to find the region in the $\mu$-$\lambda$ parameter space where
the system is unstable, a region which we call the ``footprint of the
instability.''

This footprint is not quite uniquely defined. For example, in
Fig.~\ref{fig:contour1} the region to the right of the bimodal
instability is perfectly stable in the sense that $\kappa=0$, whereas
in the region between MZA and bimodal instability, $\kappa$ is never
exactly zero but only becomes extremely small. Therefore, as a
specific criteria we adopt somewhat arbitrarily
$\kappa/\omega_0>1/100$ as a stability condition. The natural
dimension for $\kappa$ is the vacuum oscillation frequency which in
our case is of order $1~{\rm km}^{-1}$. Therefore, our instability
criteria corresponds to one $e$-folding of growth on a radial distance
of around 100~km.  Choosing a smaller $\kappa$ criteria would imply a
larger radial distance for one $e$-folding of growth.  Notice also
that the exact locus of the limiting $\kappa$ contour depends somewhat
on the exactly chosen neutrino energy distribution which here was
taken to be monochromatic. Also notice that the region nominally found
unstable in this sense may not be unstable enough to lead to any
appreciable growth in realistic situations. Our somewhat arbitrary
stability criteria is probably conservative in this sense. In any
case, we here only study the differential effect of modifying the
$\epsilon$ value so that our exact choice of stability criteria is not
crucial for our discussion.

In Fig.~\ref{fig:footsix} we show the instability footprints of
Figs.~\ref{fig:contour1} and~\ref{fig:contour2}, essentially providing
the same information as the contour plots, but now on a logarithmic
scale encompassing the range of parameters relevant in the SN context.

For inverted mass ordering and $\epsilon\agt 0$, the upper right panel
(bimodal instability) applies. For large $\mu$ and large $\lambda$, the red
instability range ($\epsilon=2/3$) is a narrow band. As shown in
Appendix~\ref{sec:explicit}, the asymptotic behavior of the instability strip
is $\lambda\propto\mu$ except for a logarithmic correction. The curves
delimiting the asymptotic footprint are provided explicitly
Appendix~\ref{sec:explicit}. For $\epsilon=0$, the asymptotic instability
region is below the large-$\epsilon$ case, but in the approximate range
$1\alt\mu\alt 100$, reducing $\epsilon$ creates an instability region above
the large-$\epsilon$ footprint. For the most part, however, reducing
$\epsilon$ extends the unstable region to the lower right of the original
footprint. We will see that this region is not of interest in the SN context.

For normal mass ordering, the panels with $\mu,\lambda<0$ apply, i.e., the
lower MZA and MAA panels. The MZA footprint is covered by the MAA one, i.e.,
in practice the MAA instability is the only one relevant in the SN context.
We see in the lower left panel that the MAA footprints are qualitatively
similar to bimodal ones. Here reducing $\epsilon$ has the clear effect of
shifting the instability region to an area between the red footprint and the
horizontal axis, which makes it ``less dangerous'' in the SN context.

\subsection{Supernova context}

\begin{figure}
\includegraphics[angle=0,width=1.0\columnwidth]{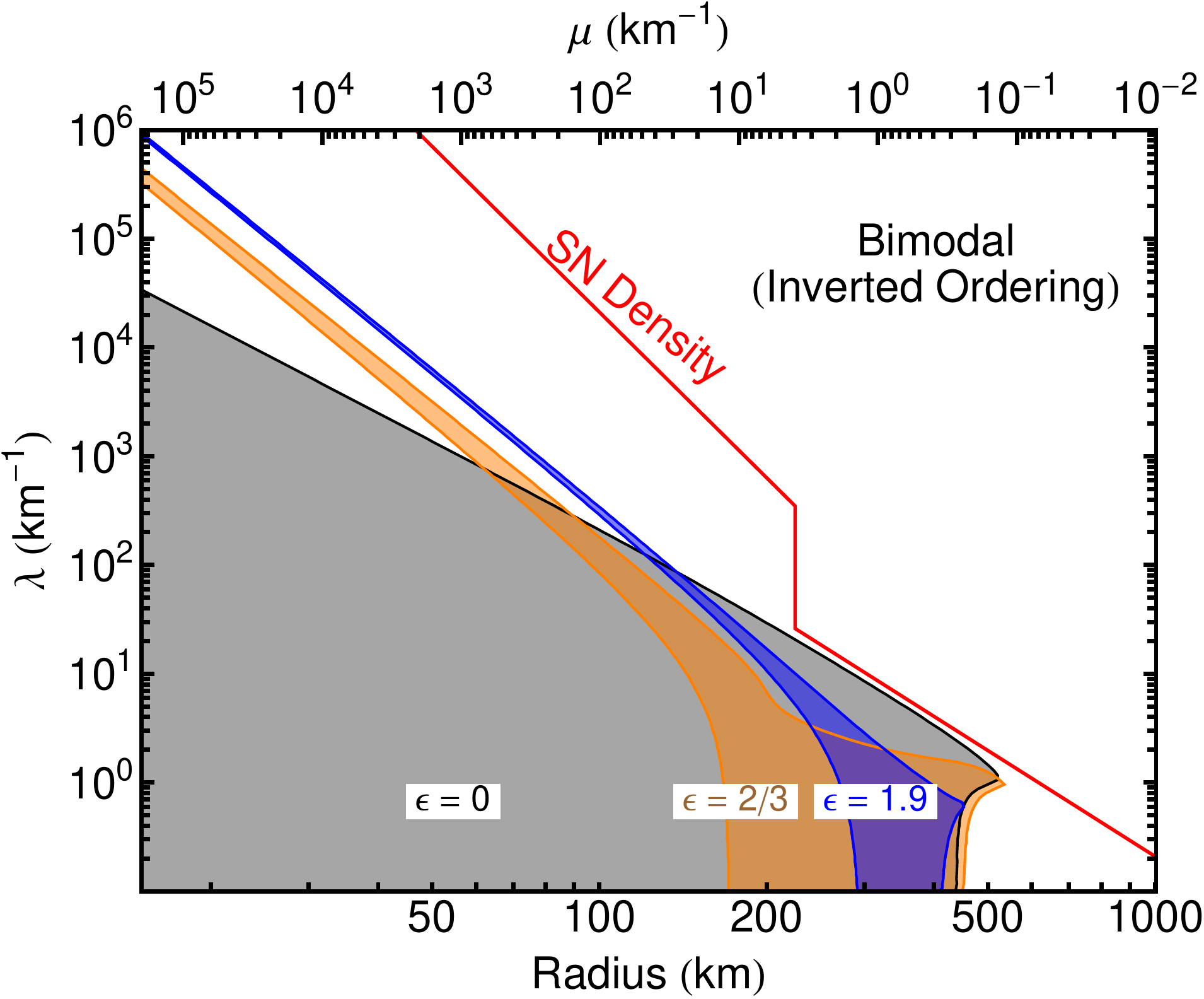}
\caption{Bimodal instability regions for the indicated
$\epsilon$ values. The $\epsilon=2/3$ and $0$ footprints are the same as those
shown in the upper right panel of Fig.~\ref{fig:footsix}. We here also show
the equivalent radius coordinate on the horizontal axis as well as a
representative SN density profile inspired by our LESA models. The steep
drop in density at around 200~km is the shock front.
 \label{fig:stab-IO}}
\vskip12pt
\includegraphics[angle=0,width=1.0\columnwidth]{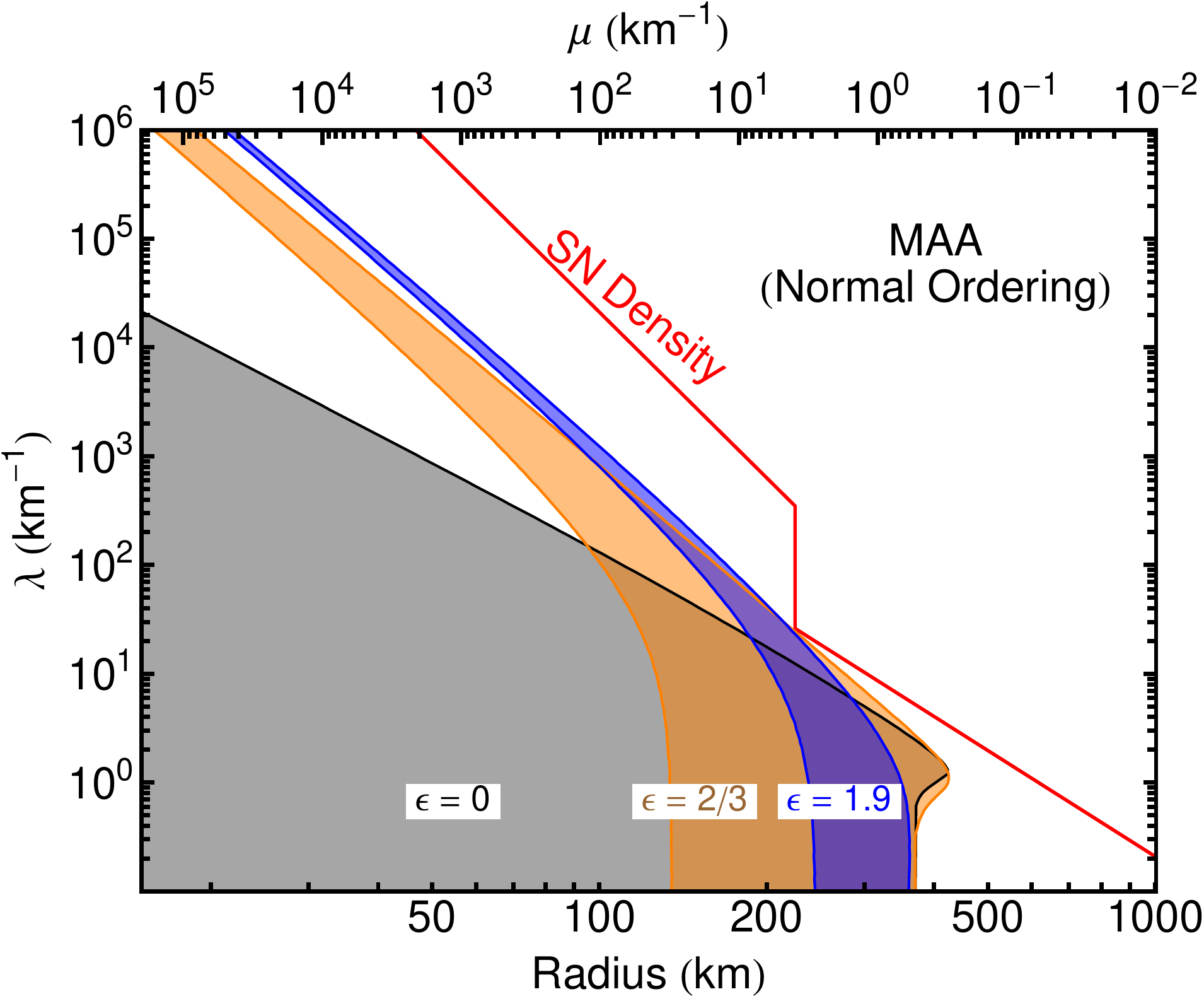}
\caption{Same as Fig.~\ref{fig:stab-IO}, now for the MAA instability region,
corresponding to the lower left panel of Fig.~\ref{fig:footsix}.
 \label{fig:stab-NO}}
\end{figure}

The findings of the previous section become more explicit if we focus
on the physical panels of Fig.~\ref{fig:footsix}, i.e., the upper
right and the lower left, and show the same information where $\mu$
and $\lambda$ are both plotted as positive variables and moreover,
where we transform the $\mu$ parameter to radial distance in a SN
model. In Figs.~\ref{fig:stab-IO} and~\ref{fig:stab-NO} we show the
bimodal and MAA instability regions and also show a representative SN
density profile inspired by our numerical LESA model.  Besides the
footprints already shown in Fig.~\ref{fig:footsix}, we now also
include the case $\epsilon=1.9$ near to the largest possible value of
$2$. As $\epsilon\to 2$, the footprint becomes an ever more
narrow sliver of parameters. Near-maximal $\epsilon$ values are
relevant in the maximum-lepton number flux direction.

In agreement with the previous literature we find that the most ``dangerous''
region with regard to a possible instability is at the shock wave where the
density drops by a large factor, corresponding to a much reduced multi-angle
matter effect. Coincidentally, in our shown example of a SN density profile, it
never intersects the instability region, but of course small changes in
chosen parameters or shifts in the density profile will lead to an unstable
region just outside of the shock wave. For a similar result see, for
example, Fig.~2 of Ref.~\cite{Raffelt:2013rqa}.

We also confirm previous findings that the MAA instability region tends to be
nearer to the SN density profile. Of course, which of the two instabilities
applies depends on the neutrino mass ordering chosen by Nature. In the MAA
case, reducing $\epsilon$ has the effect of shifting the unstable region away
from the SN density profile. In the bimodal case, the effect is not
monotonic, but typically the $\epsilon=0$ case tends to be ``most dangerous''
in the shock-wave region and comparable to the large-$\epsilon$ case
of the MAA case.

\section{Conclusions}
\label{sec:conclusions}

Motivated by the LESA effect, we have studied a simplified example of
the impact of an unusually small lepton-number flux on the interplay
between self-induced neutrino flavor conversion and its multi-angle
matter suppression. The instability regions are strongly enlarged into
regions of small matter density and large neutrino density as perhaps
expected. However, the stability plots in Figs.~\ref{fig:stab-IO} and
\ref{fig:stab-NO} show that for the most part this is a region which
does not correspond to a realistic SN density profile. Lowering the
lepton-number flux by itself does not have the effect of leading to
self-induced flavor conversion in regions below the shock front.

The main point of our study was to explore the direction of the effect
of lowering the lepton-number flux. According to our simplified models
even strong modifications of the asymmetry parameter do not have a
dramatic impact on the stability question along realistic SN density
profiles. For IO, where the bimodal instability applies, a small
$\epsilon$ value can have the effect of creating an unstable region
beyond the shock-wave radius. For NO, where the MAA instability
applies, the largest $\epsilon$ values provide the ``most dangerous''
instability regions.

Overall it is primarily the multi-angle matter effect which controls
the instability region for densities relevant along the SN
profile. For practical SN simulations with regard to the explosion
mechanism, it appears that unusually small $\epsilon$ values do not
require a reconsideration of neutrino flavor evolution for distances
below the shock front. Of course, flavor conversion will be important
beyond of the shock front, certainly at large distances by the MSW
effect, and perhaps in some cases directly outside of the shock front
by self-induced flavor conversion. Such effects would be important for
the interpretation of the neutrino signal from the next nearby SN, but
would not directly affect the explosion dynamics.

Our stability analysis was somewhat schematic with the purpose of exploring the
general direction of what an unusually small lepton-number flux caused by the
LESA effect would do.  In principle, of course, in any given SN simulation
one could perform an {\em a posteriori\/} stability analysis, based on the numerical
neutrino radiation field and density profile, to verify that it was justified
to ignore flavor conversion. In 3D simulations, the local neutrino radiation
field will not be axially symmetric, not even in the LESA dipole directions.
Therefore, the stability analysis has to be based on neutrino fluxes with a
non-trivial azimuth-angle distribution. This is not conceptually difficult,
but such cases have not yet been explored in practice and await systematic
study. In principle, one could eventually develop a numerical tool that could
flag possible flavor instabilities as the numerical simulation progresses. Of
course, should cases be found where flavor conversion below the shock front
occurs after all, a completely new challenge has to be faced in SN modeling.

We stress, however, that not finding flavor-conversion instabilities in such
an approach does not necessarily prove that none exist.  As explained in the
introduction, we have ignored a number of new issues that have emerged in the
recent literature on neutrino-neutrino refractive effects. These topics
should be sorted out before worrying further about self-induced flavor
conversion in practical SN simulations.  In particular, the multi-angle
matter effect which provides the stabilizing ingredient in the present
context relies on the assumption that the neutrino radiation field and its
flavor properties vary only along the radial direction. This is an imposed
symmetry assumption which can hide unstable solutions that might exist
otherwise. Likewise, the assumption of a purely stationary solution, allowing
us to treat the problem in the form of an ordinary (rather than partial)
differential equation has never been strictly justified.

The core-collapse SN explosion mechanism is one of the few physical phenomena
where neutrinos play a dominant dynamical role and where the flavor
dependence of the fluxes matters. Yet flavor conversion, in spite of large
mixing angles, does not seem to figure at all for SN dynamics due to matter
suppression of flavor oscillations, i.e., because in dense matter,
propagation and interaction eigenstates are almost the same. However, a final
verdict on the role of active-active flavor conversion for SN dynamics
requires more theoretical work to fully appreciate the role of possible
flavor instabilities in the interacting neutrino field.

\section*{Acknowledgments}

This work was partly supported by the Deutsche Forschungsgemeinschaft under
Grant No.\ EXC-153 (Cluster of Excellence ‘‘Origin and Structure of the
Universe’’) and by the European Union under Grant No.\ PITN-GA-2011-289442
(FP7 Initial Training Network ‘‘Invisibles’’). S.C.\ acknowledges support
from the European Union through a Marie Curie Fellowship, Grant No.\
PIIF-GA-2011-299861. B.M.\ has been supported by the Australian Research Council
through a Discovery Early Career Researcher Award (Grant No.\ DE150101145).

\appendix

\section{Explicit eigenvalue equation}
\label{sec:explicit}

\subsection{Explicit integrals}

In order to solve the eigenvalue equation we need the integrals defined in
Eq.~(\ref{eq:In-definition}). In our simplified model, they take on the form
\begin{equation}\label{eq:Kn-definition-1}
I_n=\frac{\mu}{\bar\lambda}\int d\omega\,h_\epsilon(\omega)\,
K_n\left(\frac{\omega-\Omega}{\bar\lambda}\right)
\end{equation}
where $\bar\lambda=\lambda+\epsilon\mu$ and
\begin{equation}\label{eq:Kn-definition-2}
K_n(w)=\int_0^1 du\,\frac{u^n}{u+w}
\end{equation}
and $w=(\omega-\Omega)/(\lambda+\epsilon\mu)$. We find explicitly
\begin{subequations}
\begin{eqnarray}
K_0(w)&=&\log\left(\frac{1+w}{w}\right)\,,\\
K_1(w)&=&1-w\,\log\left(\frac{1+w}{w}\right)\,,\\
K_2(w)&=&\frac{1}{2}-w+w^2\log\left(\frac{1+w}{w}\right)\,.
\end{eqnarray}
\end{subequations}
As a next step we adopt the monochromatic spectrum Eq.~(\ref{eq:monospectrum})
and express all frequencies in units of $\omega_0$, i.e., $\mu$ is understood to
mean $\mu/\omega_0$ and so forth. With $\bar\lambda=\lambda+\epsilon\mu$
we find explicitly
\begin{widetext}
\begin{subequations}
\begin{eqnarray}
\kern-1em
I_0&=&\frac{\mu}{\bar\lambda}\,\,\left[\left(1+\frac{\epsilon}{2}\right)\log\left(1-\frac{\bar\lambda}{\Omega-1}\right)
-\left(1-\frac{\epsilon}{2}\right)\log\left(1-\frac{\bar\lambda}{\Omega+1}\right)\right]\,,\\[1ex]
\kern-1em
I_1&=&\frac{\mu}{\bar\lambda^2}\left[\epsilon\bar\lambda+
\left(1+\frac{\epsilon}{2}\right)(\Omega-1)\log\left(1-\frac{\bar\lambda}{\Omega-1}\right)
-\left(1-\frac{\epsilon}{2}\right)(\Omega+1)\log\left(1-\frac{\bar\lambda}{\Omega+1}\right)\right]\,,\\[1ex]
\kern-1em
I_2&=&\frac{\mu}{\bar\lambda^3}\left[\bar\lambda\,\left(\frac{\epsilon\bar\lambda}{2}+\epsilon\Omega-2\right)+
\left(1+\frac{\epsilon}{2}\right)(\Omega-1)^2\log\left(1-\frac{\bar\lambda}{\Omega-1}\right)
-\left(1-\frac{\epsilon}{2}\right)(\Omega+1)^2\log\left(1-\frac{\bar\lambda}{\Omega+1}\right)\right]\,.
\end{eqnarray}
\end{subequations}
\end{widetext}

\subsection{Vanishing effective matter density (\boldmath{$\bar\lambda\to 0$})}

In the absence of matter effects, where $\bar\lambda=\lambda+\epsilon\mu\to
0$, these expressions simplify to
\begin{equation}
I_0=\mu\,\frac{2+\epsilon\Omega}{1-\Omega^2}\,,\quad
I_1=\frac{1}{2}\,I_0\,,\quad\hbox{and}\quad
I_2=\frac{1}{3}\,I_0\,.
\end{equation}

In this case the eigenvalue equations can be solved explicitly,
\begin{subequations}
\begin{eqnarray}
\kern-3em
\Omega_{\rm bimodal}&=&-\frac{2\sqrt3+3}{12}\,\epsilon\mu
\nonumber\\
&\pm&\sqrt{1-\frac{2\sqrt3+3}{3}\,\mu+\frac{7+4\sqrt3}{48}\left(\epsilon\mu\right)^2}
\,,\\[1ex]
\kern-3em
\Omega_{\rm MZA}&=&+\frac{2\sqrt3-3}{12}\,\epsilon\mu
\nonumber\\
&\pm&\sqrt{1+\frac{2\sqrt3-3}{3}\,\mu+\frac{7-4\sqrt3}{48}\left(\epsilon\mu\right)^2}
\,,\\[1ex]
\kern-3em
\Omega_{\rm MAA}&=&\frac{\epsilon\mu}{4}\pm
\sqrt{1+\mu+\left(\frac{\epsilon\mu}{4}\right)^2}
\,.
\end{eqnarray}
\end{subequations}
For these solutions to have a nonvanishing imaginary part, the bimodal
solution requires $\mu>0$, whereas the two others require $\mu<0$. For all
three cases, one finds that the maximum growth rate to be
\begin{equation}
\kappa_{\rm max}=\sqrt{(2/\epsilon)^2-1}\,,
\end{equation}
which of course only applies on the locus $\bar\lambda=0$, i.e., on the line
$\lambda=-\epsilon\mu$. In other regions of the $\mu$-$\lambda$ parameter
space, $\kappa_{\rm max}$ could be larger. Notice that in our convention,
$-2\leq\epsilon\leq+2$ so that $\kappa_{\rm max}$ is indeed real and
positive. For small $\epsilon$, we have $\kappa_{\rm max}\approx
|2/\epsilon|$. Recalling that $\kappa$ is given in units of the vacuum
oscillation frequency $\omega_0$ we find once more that the maximum growth
rate is essentially identical with the vacuum oscillation frequency times a
numerical factor depending on~$\epsilon$.

The situation changes in the absence of any $\nu_e$-$\bar\nu_e$ asymmetry
where $\epsilon=0$. In this case the imaginary parts of all three solutions
grow without limit for $|\mu|\to\infty$.  For $|\mu|=400$, the edge of the
plotting region in Fig.~\ref{fig:contour2}, and for $\lambda=0$ we find
$\kappa_{\rm bimodal}=29.34$, $\kappa_{\rm MZA}=7.93$, and  $\kappa_{\rm
MAA}=20.02$, in agreement with the contours in Fig.~\ref{fig:contour2}.

\subsection{Large effective matter density (\boldmath{$\bar\lambda\to\pm\infty$})}

In the opposite limit of very large effective matter density
$\bar\lambda\to\infty$ we may also gain considerable insight by
analytic techniques. To this end we observe that unstable solutions,
for large $\lambda$ values, also require large $\mu$
values. Therefore, to expand in powers of $1/\bar\lambda$, we express
$\mu=m\,\bar\lambda$, where $m$ is a dimensionless parameter. Of
course, at this stage this is only an ansatz because $\mu$ could also
scale with another power of $\bar\lambda$.

Considering first $\bar\lambda>0$ we assume that the real part of
$\Omega$ does not become large, i.e., that $|\Omega|$ can be
considered small relative to $\bar\lambda$ or $\mu$. Under these
conditions, to lowest order in $1/\bar\lambda$ the first block of the
eigenvalue equation (\ref{eq:stabilityconidtion}) becomes
\begin{equation}\label{eq:logequation}
\frac{(2+\epsilon)\log(1-\Omega)-(2-\epsilon)\log(-1-\Omega)}{2\epsilon}=a\,,
\end{equation}
where
\begin{equation}\label{eq:am1}
a=\log\left(\bar\lambda\right)-\frac{2\,(m-1)^2}{m^2}\,.
\end{equation}
The second block, yielding the MAA instability, provides to lowest
order
\begin{equation}
m=-1\,.
\end{equation}
Therefore, to lowest order in $1/\bar\lambda$, any solution, stable or
unstable, requires $\mu=-\bar\lambda$. Therefore, with
$\bar\lambda=\lambda+\epsilon\mu$ we find
\begin{equation}\label{eq:asym-MAA-plus}
\mu=-\frac{\lambda}{1+\epsilon}\,.
\end{equation}
This results corresponds to the very thin footprint in the upper left panel
of Fig.~\ref{fig:footsix}.

Next we turn to $\bar\lambda\to-\infty$. We notice without further
discussion that in this case the real part of $\Omega$ becomes of
order $\bar\lambda$ so that we write
$\Omega\to\bar\lambda+\Omega$. Performing now the expansion, we find
for both blocks of Eq.~(\ref{eq:stabilityconidtion}) a result of the
form of Eq.~(\ref{eq:logequation}), but with modified expressions
\begin{subequations}\label{eq:am2}
\begin{eqnarray}
a&=&\log\left(-\bar\lambda\right)-\frac{2\,(m-1)^2}{m(m-4)}
\,,\\
a&=&\log\left(-\bar\lambda\right)-\frac{m+1}{m}
\,,
\end{eqnarray}
\end{subequations}
which appear on the right-hand side.

Equation~(\ref{eq:logequation}) can be solved explicitly for our
benchmark example $\epsilon=2/3$ where it is essentially a quadratic
equation. In this case one finds
\begin{equation}
\Omega=\frac{2-e^{a}\pm e^{a/2}\sqrt{e^a-8}}{2}\,.
\end{equation}
This solution has a nonvanishing imaginary part for $-\infty<a<\log(8)$.
However, for $a\ll -1$ the imaginary part (the growth rate) is exponentially
suppressed. The formal criteria $\kappa>1/100$ leads to the requirement
$a>\log(4-\sqrt{39999}/50)$. Therefore, the system is unstable for
$a_1<a<a_2$, where
\begin{subequations}
\begin{eqnarray}
a_1&=&\log(4-\sqrt{39999}/50)\approx-9.90\,,
\\
a_2&=&\log(8)\approx+2.08\,,
\end{eqnarray}
\end{subequations}
assuming our nominal requirement $\kappa>1/100$.

To make these results more explicit for the various instabilities, we first
solve Eq.~(\ref{eq:am1}), corresponding to $\bar\lambda\to+\infty$. We find
\begin{eqnarray}\label{eq:asym-bim-plus}
\mu=\frac{\bar\lambda}{1\pm\sqrt{(\log\bar\lambda-a)/2}}\,.
\end{eqnarray}
Together with $\lambda=\bar\lambda-\epsilon\mu$ and with the explicit values
for $a_1$ and $a_2$ we can find the limiting contours in the
$\mu$-$\lambda$-plane as a parametric plot depending on the
variable~$\bar\lambda$.

Likewise, we may solve Eq.~(\ref{eq:am2}), corresponding to
$\bar\lambda\to-\infty$. Here we find
\begin{subequations}
\begin{eqnarray}\label{eq:asym-minus}
\mu&=&+\frac{\bar\lambda}{1+A\pm\sqrt{\frac{3}{2}A+A^2}}\,,
\\
\mu&=&-\frac{\bar\lambda}{1+A}\,,
\end{eqnarray}
\end{subequations}
where
\begin{equation}
A=a-\log\bar\lambda\,.
\end{equation}
Again we can find the limiting contours with $a_1$ and $a_2$ in the
$\mu$-$\lambda$-plane as a parametric plot depending on the variable
$\bar\lambda$.


\end{document}